\documentclass[aip,jcp,reprint,longbibliography]{revtex4-1}
\usepackage{epsfig}
\usepackage{color}
\usepackage{amsmath}
\usepackage{amsfonts}
\usepackage{natbib}
\usepackage{notes2bib}

\newcommand{\degree}{^{\circ} }

\newcommand{\avg}[1]{\left<#1\right>}

\newcommand{\brac}[1]{\left[#1\right]}

\newcommand{\para}[1]{\left(#1\right)}

\newcommand{\kT}{\ensuremath{k_{\rm B}T}}

\newcommand{\nb}{\ensuremath{\mathbf{n}}}

\begin{document}

%\preprint{}

%Title of paper - Just an idea for now, we'll probably need a better one...
\title{Molecular Structure and Rotational Dynamics in the Acetonitrile:Acetylene (1:2) Plastic Co-Crystal at Titan Conditions}


\author{Atul C. Thakur}
\author{Richard C. Remsing}
\email[]{rick.remsing@rutgers.edu}
\affiliation{Department of Chemistry and Chemical Biology, Rutgers University, Piscataway, NJ 08854}

%\thanks{}


%\date{\today}

\begin{abstract}
The surface of Saturn's moon Titan is coated with small molecule organic solids termed cryominerals. 
Cryominerals play an analogous role to minerals on Earth in Titan's surface geology and geochemistry. 
To develop a predictive understanding of Titan's surface geochemistry, we need to characterize the structure and dynamics of cryominerals at the molecular scale. 
We use ab initio molecular dynamics simulations to quantify the structure and dynamics of the acetonitrile:acetylene (1:2) co-crystal at Titan surface conditions.
We suggest that acetonitrile:acetylene is in a plastic phase, in which acetonitrile molecules are dynamically disordered about the N-C-C axis on sub-picosecond timescales, and that this rotational, plastic disorder persists at least to temperatures of 30~K. 
We anticipate that many cryominerals may have plastic phases at or near Titan surface conditions, and understanding this disorder will be crucial to predicting chemistry on Titan's surface.
\end{abstract}

%\keywords{}

\maketitle

\raggedbottom

%**********************************************************************************%
\section{Introduction}
The surface of Titan, Saturn's largest moon, is coated with solid organics. 
These organic molecules are produced through photochemistry in Titan's thick, chemically rich upper atmosphere.
The cold temperature of 94~K leads to their condensation onto Titan's surface~\cite{lorenz2006sand,singh2016acetylene,horst2017titan,lunine2020astrobiology,mackenzie2021titan,nixon2024composition}.
As a result, understanding Titan's geological features and its surface geochemistry necessarily involves understanding these organic solids. 
These small molecule organic solids are expected to play a role analogous to the minerals on Earth, and consequently, Titan's organic crystals have been termed cryominerals~\cite{maynard2018prospects,cable2021titan}.
The emerging field of Titan cryomineralogy has resulted in the prediction and discovery of many potential cryominerals~\cite{maynard2018prospects,cable2021titan,mcconville2020peritectic}, whose physical properties need to be understood to create a thorough understanding of Titan's geochemistry and the evolution of Titan's surface. 
Organic solids are typically held together by weak interactions.
As a result, they often exhibit rich phase diagrams with varying ordered phases but also can exhibit phases with disorder.
Of particular interest are plastic crystal (or rotator) phases, where molecules are translationally ordered but at least one component of the solid is orientationally disordered.
The orientational disorder present in plastic phases alters thermodynamic and mechanical properties relative to the ordered crystal~\cite{klein1990simulation,klein1985computer,lynden1994translation} 
and can also impact reactivity~\cite{tran2005photochemical,gavezzotti1982crystal,shalaev2002effect} and transport~\cite{zhang2022exploiting, dhattarwal2024electronic}, such that characterizing the prevalence and physical properties of plastic crystals is crucial to understanding Titan's geochemistry.
We previously predicted that the acetylene:ammonia (1:1) co-crystal is in a plastic phase at Titan surface temperatures and thoroughly characterized its molecular structure and dynamics~\cite{thakur2023molecular,thakur2024nuclear}. 
In this work, we predict that another cryomineral, the acetonitrile:acetylene (1:2) co-crystal~\cite{cable2020properties}, is a plastic crystal at Titan conditions.
We use ab initio molecular dynamics (AIMD) simulations to model the acetonitrile:acetylene (1:2) co-crystal~\cite{cable2020properties} and characterize its molecular structure and dynamics. 
We predict that the acetonitrile molecules within the co-crystal exhibit dynamic orientational disorder on sub-picosecond timescales. 
The presence of disorder results from the ability of thermal fluctuations to easily disrupt weak interactions between the acetonitrile CH$_3$ group and surrounding molecules. 
%

%**********************************************************************************%

\section{Simulation Details}
We performed density functional theory (DFT)-based Born-Oppenheimer molecular dynamics (BOMD) simulations of the acetonitrile:acetylene (1:2) co-crystal utilizing the CP2K software package~\cite{vandevondele2003efficient,hutter2014cp2k,CP2K}. 
The QUICKSTEP electronic structure module, a component of the CP2K software, was utilized for precise and efficient simulations within the framework of DFT.
The CP2K QUICKSTEP module employs a mixed Gaussian and plane wave (GPW) basis enhancing algorithmic efficiency for both electrostatics and exchange-correlation energies~\cite{lippert1997hybrid}.
We utilized molecularly optimized (MOLOPT) G\"{o}decker-Teter-Hutter (GTH) triple-$\zeta$ single polarization (TZVP-MOLOPT-GTH) basis sets implemented in CP2K for representing orbitals~\cite{vandevondele2007gaussian}.
The electron density was accurately represented by determining the grid spacing via a plane wave 
cutoff of 500~Ry. 
Core electrons were represented using GTH pseudopotentials~\cite{hartwigsen1998relativistic,goedecker1996separable,krack2005pseudopotentials}.
We adopted the Perdew-Burke-Ernzerhof (PBE) exchange-correlation (XC) functional~\cite{perdew1996generalized} supplemented with Grimme's D3 corrections for long-range dispersion interactions~\cite{grimme2010consistent,grimme2011effect}.

Our simulations used the experimentally-determined crystal structure of the acetonitrile:acetylene (1:2) co-crystal deposited in the Cambridge Structural Database with ID 735040~\cite{kirchner2010co}. 
A $2\times 2 \times 2$ supercell of this experimental crystal structure containing 448 atoms was created using CP2K.
We carried out BOMD simulations of the co-crystal at two temperatures, 30~K and 90~K.  
The equations of motion were propagated using the velocity-Verlet scheme in CP2K with a timestep of 1~fs. 
In each BOMD run, the system was equilibrated at a given temperature for at least 3~ps in the canonical (NVT) ensemble using the canonical stochastic velocity rescaling (CSVR) thermostat~\cite{bussi2007canonical}, before performing production runs in the microcanonical (NVE) ensemble.
The 90~K simulation resulted in a trajectory approximately 58~ps in length, while the 30~K simulation resulted in a trajectory length of approximately 24~ps.  
%

%------------------------------------------------------------------------------------------------------------------------------------------------------------------
\begin{figure}[tb]
\begin{center}
\includegraphics[width=0.495\textwidth]{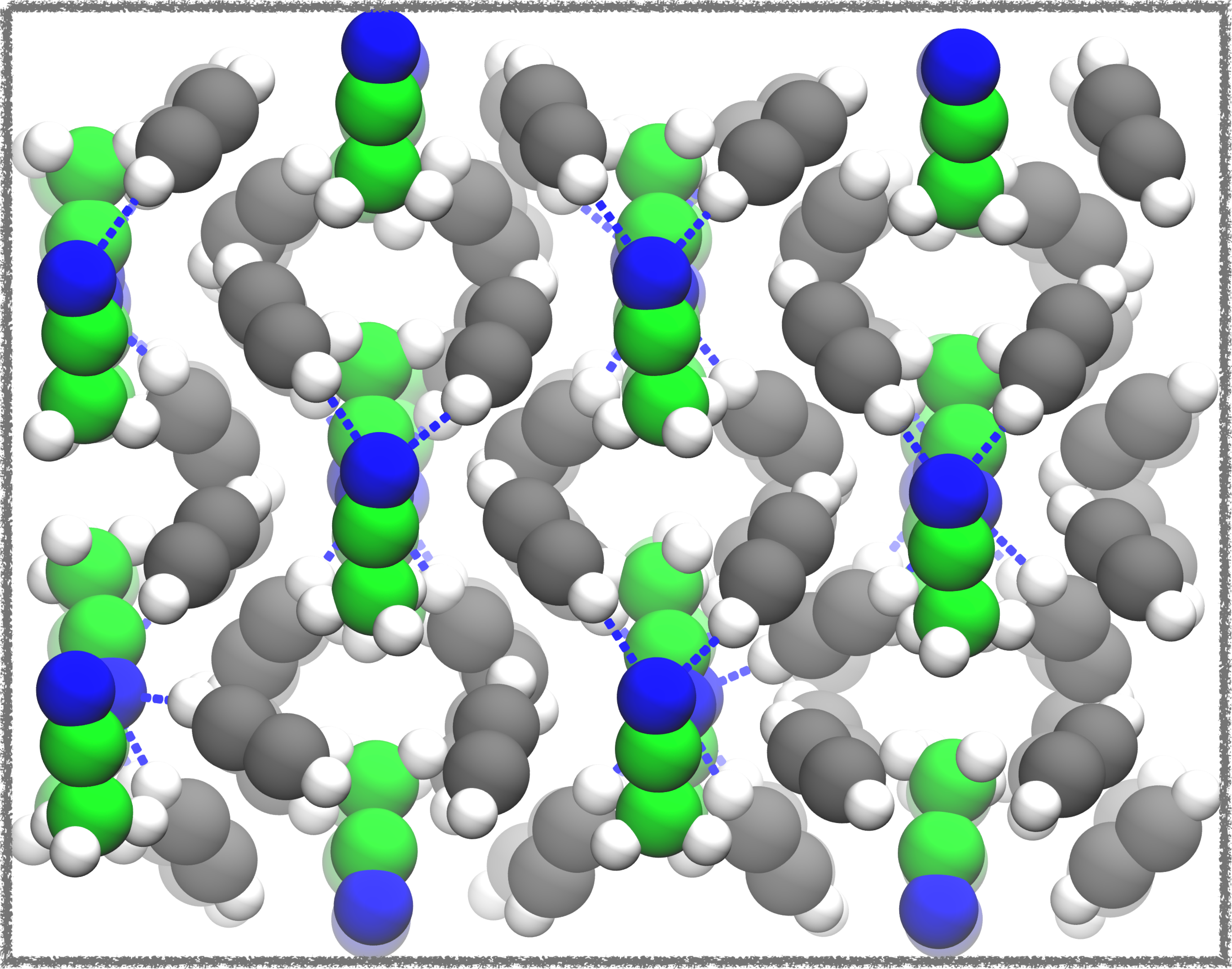}
\end{center}
\caption
{
Snapshot of the acetonitrile:acetylene (1:2) co-crystal from BOMD simulations of the co-crystal at 90~K. 
The blue dotted lines depict CH$\cdots$N hydrogen bonds donated from the acetylene to the acetonitrile molecules.
The acetylene carbons are gray and the acetonitrile carbons are drawn using green color for clarity.
The nitrogen atoms of acetonitrile are colored blue, while all hydrogen in the system are colored white.
The snapshot shows the view from the crystal $z$ axis.
}
\label{fig:snap}
\end{figure}
%------------------------------------------------------------------------------------------------------------------------------------------------------------------

%**********************************************************************************%     
\section{Results and Discussion}

%**********************************************************************************%
\subsection{Directional Interactions in Acetonitrile:Acetylene}

We first analyze structural correlations due to directional interactions within the acetonitrile:acetylene co-crystal, whose structure is shown in Fig.~\ref{fig:snap}.
Specifically, we examine spatial correlations indicative of three types of hydrogen bonds that could form in the co-crystal: (1) acetylene C-H$\cdots$N, (2) acetonitrile C-H$\cdots$N, and (3) acetonitrile C-H$\cdots\pi$ of acetylene, where the $\pi$ system is spatially represented by the mid-point of the acetylene C-C triple bond.
We quantify these correlations through the free energy surface
\begin{equation}
\Delta A(\beta,R) = -\kT \ln P(\beta,R),
\end{equation}
where $\kT$ is the product of Boltzmann's constant and the temperature, $\beta$ is the acceptor-donor-hydrogen angle, $R$ is the donor-acceptor distance, and $P(\beta,R)$ is a two-dimensional probability distribution~\cite{kumar2007hydrogen}.
%

%------------------------------------------------------------------------------------------------------------------------------------------------------------------
\begin{figure*}[tb]
\begin{center}
\includegraphics[width=0.96\textwidth]{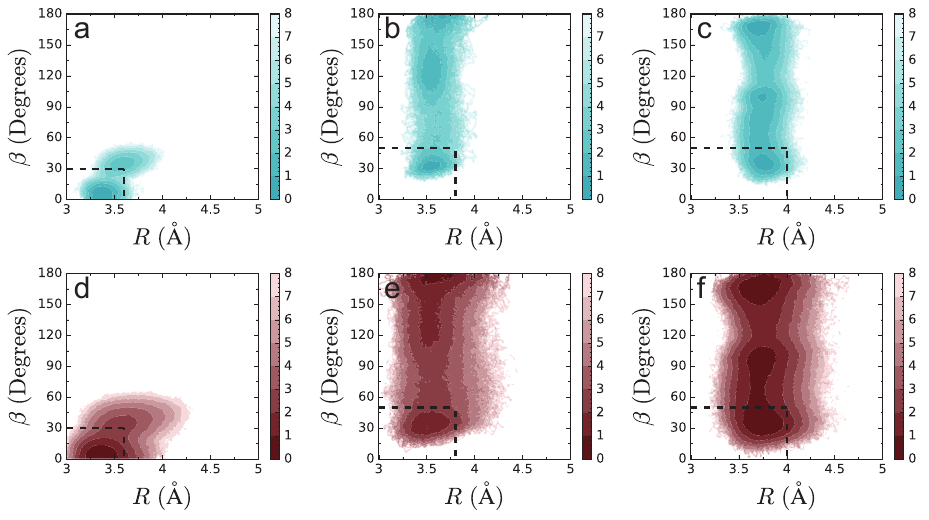}
\end{center}
\caption
{
Free energy surfaces, $\Delta A(\beta,R)/\kT$, as a function of acceptor-donor-hydrogen angle, $\beta$, and donor-acceptor distance, $R$ at (a-c) 30~K and (d-f) 90~K. 
Free energies are shown for (a,d) acetylene C-H$\cdots$N, (b,e) acetonitrile C-H$\cdots$N, and (c,f) acetonitrile C-H$\cdots\pi$ interactions.
}
\label{fig:fes}
\end{figure*}
%------------------------------------------------------------------------------------------------------------------------------------------------------------------

%%
%
We find that only acetylene C-H$\cdots$N hydrogen bonds are formed in the acetonitrile:acetylene co-crystal at both temperatures of interest, Fig.~\ref{fig:fes}a,d, because only these free energies have a minimum consistent with a linear H-bond arrangement at short donor-acceptor distances.
At 30~K, $\Delta A(\beta,R)$ exhibits a minimum near 3.4~\AA \ and $0\degree$ consistent with H-bonding, and a local minimum near 3.6~\AA \ and approximately $40\degree$. 
The local minimum arises from the acetonitrile molecule's third nearest acetylene neighbor, to which it is not H-bonded at 30~K.
At 90~K, the local minimum smears out into a shoulder due to increased thermal fluctuations in the crystal.
In this case, the third nearest acetylene neighbor can occasionally donate an additional hydrogen bond to the acetonitrile molecule, using a geometric definition of a hydrogen bond~\cite{luzar1996hydrogen,kumar2007hydrogen} indicated by the dashed lines in Fig.~\ref{fig:fes}a,d.
The other two types of correlations, acetonitrile C-H$\cdots$N and acetonitrile C-H$\cdots\pi$ of acetylene, rarely result in hydrogen bonds, based on the above geometric criterion.
Instead, we observe weak directional interactions at larger angles and distances, Fig.~\ref{fig:fes}, which can be broken with free energy changes on the order of the thermal energy, $\kT$.
Despite their relatively weak strength, we can still define geometric criteria for the existence of these interactions, indicated by the dashed lines in Fig.~\ref{fig:fes}b,e for acetonitrile C-H$\cdots$N interactions and Fig.~\ref{fig:fes}c,f for acetonitrile C-H$\cdots\pi$ interactions.
We use these definitions later to quantify the dynamics associated with the lifetime of these interactions.
In summary, we find that the acetonitrile:acetylene co-crystal is held together by acetylene C-H$\cdots$N hydrogen bonds and weak but directional acetonitrile C-H$\cdots$N and acetonitrile C-H$\cdots\pi$ interactions, in addition to non-specific van der Waals interactions.

%**********************************************************************************%
\subsection{Disordered Orientational Structure}

\begin{figure}[tb]
\begin{center}
\includegraphics[width=0.48\textwidth]{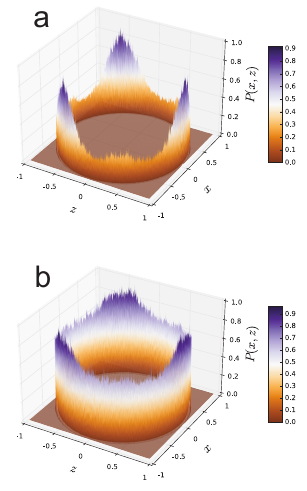}
\end{center}
\caption
{
Probability distributions of the C-H bond vectors of acetonitrile at temperatures of (a) 30~K and (b) 90~K. 
Both the colorbar and $P(x, z)$ axes indicate the probability of an orientation projected onto the $xz$ plane. 
}
\label{fig:os}
\end{figure}

The weak interactions in the co-crystal can be readily disrupted by thermal fluctuations to create disorder.
We quantify disorder in the acetonitrile:acetylene by constructing probability distributions of the orientation of the acetonitrile C-H bond vectors. 
The orientations of the C-H bonds are not directly along a crystallographic plane.
Therefore, we quantify the C-H bond orientation by projecting the C-H bond vectors onto the instantaneous plane formed by the three hydrogen atoms of an acetonitrile's CH$_3$ group, and then we rotate this projection onto the global $xz$-plane. 
If the system is in a purely crystalline phase, there is no orientational disorder, and peaks in the distribution will be disconnected~\cite{thakur2023molecular}; regions of zero probability will separate the peaks in the orientational distribution.
If the system is in a plastic crystal phase, orientational disorder of the C-H bonds will lead to non-zero probability connecting the peaks in the distribution.
We find that the acetonitrile:acetylene co-crystal at 90~K follows the latter behavior, suggesting that it is in a plastic crystal phase at Titan's surface temperature, Fig.~\ref{fig:os}.
%

%------------------------------------------------------------------------------------------------------------------------------------------------------------------
%------------------------------------------------------------------------------------------------------------------------------------------------------------------

%%
%
The orientational probability distributions computed at temperatures of 30~K and 90~K display three broad peaks and significant probability connecting the peaks. 
The fully connected probability distributions indicate that the acetonitrile molecules are orientationally disordered at Titan's surface temperature of 90~K and that this disorder remains at the low temperature of 30~K. 
The presence of orientational disorder at 30~K contrasts previous results for the acetylene:ammonia co-crystal, which is a plastic crystal at 90~K but not at 30~K, where it is an ordered crystal.
We attribute this difference to the presence of hydrogen bonds in the acetylene:ammonia co-crystal, such that 30~K does not provide enough thermal energy to break these hydrogen bonds~\cite{thakur2023molecular,thakur2024nuclear}.
In the acetonitrile:acetylene co-crystal, there are no strong interactions between acetonitrile and acetylene that need to be disrupted for rotations to occur, and as a result, acetonitrile:acetylene can remain in the plastic phase over a much larger range of temperatures.
However, even weak directional interactions lead to regions of higher probability and preferred orientations, especially at 30~K, but these preferred orientations are separated from others by free energy scales less than $2\kT$ at 30~K and less than $\kT$ at 90~K.
We note that we modeled the nuclei classically and neglected nuclear quantum effects~\cite{markland2018nuclear}.
Nuclear quantum effects can increase orientational disorder in plastic crystals~\cite{thakur2024nuclear}, especially when the disorder involves light atoms like hydrogen.
As a result, we anticipate that a more accurate description of the co-crystal that also treats the nuclei quantum mechanically would only increase the amount of disorder, such that the prediction of an orientationally disordered acetonitrile:acetylene co-crystal at low temperatures is robust. 
%

%**********************************************************************************%
\subsection{Rotational Dynamics of Acetonitrile}
The presence of structural disorder suggests that the acetonitrile:acetylene co-crystal is either a plastic crystal or an orientational glass at Titan conditions.
In a plastic crystal phase, the disorder is dynamic and the acetonitrile molecules will dynamically rotate, whereas in an orientational glass, the acetonitrile molecules would be orientationally disordered but not actively rotating.
To discern the phase of the acetonitrile:acetylene co-crystal, we quantified the rotational dynamics of the acetonitrile molecules. 
To do so, we computed the rotational time correlation function
\begin{equation}
C_2(t) = \avg{P_2\para{\hat{\nb}(0)\cdot \hat{\nb}(t)}}
\end{equation}
for acetonitrile and acetylene, where $\hat{\nb}(t)$ is the C-H bond unit vector at time $t$ for acetonitrile or the C-C bond unit vector at time $t$ for acetylene, and $P_2(x)$ is the second-order Legendre polynomial.
This form of the rotational correlation function can in principle be extracted from Raman spectroscopy and the rotational correlation time, given by the integral of $C_2(t)$, can be measured using nuclear magnetic resonance (NMR) relaxometry~\cite{gordon1965relations,gordon1965molecular,thakur2021distributed}. 
%

%------------------------------------------------------------------------------------------------------------------------------------------------------------------
\begin{figure}[tb]
\begin{center}
\includegraphics[width=0.45\textwidth]{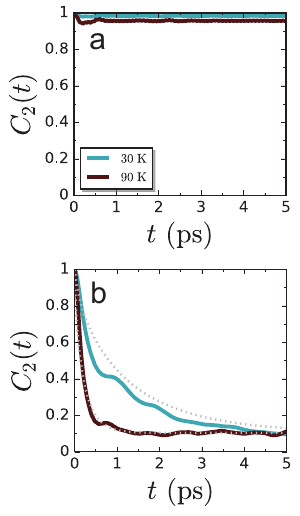}
\end{center}
\caption
{
Rotational time correlation functions for the (a) acetylene C-C bond vector and (b) acetonitrile C-H bond vector at 30~K and 90~K. 
Dashed gray lines in (b) show predictions of $C_2(t)$ using the Lipari and Szabo approximation~\cite{lipari1981pade} for free rotation within a cone with semiangle $\theta_0=63.5\degree$.
}
\label{fig:tcf}
\end{figure}
%------------------------------------------------------------------------------------------------------------------------------------------------------------------

%%
%
The rotational time correlation function, $C_2(t)$, hardly decays for acetylene molecules, Fig.~\ref{fig:tcf}a.
The lack of substantial decay indicates that acetylene molecules do not rotate.
The small decay is due to librational modes that cause small changes in the tilt of the C-C bond vector.

For the acetonitrile molecules, $C_2(t)$ decays nearly completely, Fig.~\ref{fig:tcf}b.
This decay indicates that acetonitrile molecules rotate around their $C_3$ symmetry axis.
As a result, the acetonitrile:acetylene co-crystal is a plastic crystal.
The time correlation function for acetonitrile rotations decay on sub-picosecond timescales, which indicates fast rotational motion within the plastic phase. 
We estimate the rotational correlation time at 90~K as $\tau_{90} = \int_0^\infty dt \brac{C_2(t)-C_2(\infty)} \approx0.15$~ps.
This rotational time is faster than previously predicted for ammonia rotations within the acetylene:ammonia co-crystal~\cite{thakur2024nuclear,thakur2023molecular}.
The faster acetonitrile rotation observed here can be attributed to the weaker interactions between the acetonitrile and acetylene molecules, as compared to the hydrogen bonding interactions between ammonia and acetylene. 
The rotational time correlation function for acetonitrile also decays at 30~K, albeit on a longer timescale of $\tau_{30}\approx 0.90$~ps, indicating that acetonitrile:acetylene is in a plastic phase even at low temperatures. 
We note that $C_2(t)$ for 30~K and 90~K do not fully decay but plateau at a value of
$\avg{P_2\para{\cos\theta}}= \para{3\avg{\cos^2\theta}-1}/2 \approx 0.1$,
where $\theta$ is the angle formed by $\hat{\nb}$ and the static averaged acetonitrile $C_3$ symmetry axis~\cite{lipari1980effect,lipari1981pade}. 
This is because rotations around acetonitrile's long axis cannot completely sample the space of rotations.
The acetonitrile rotations are restricted because the CH$_3$ group remains pointed in the same direction within the solid.
To highlight this restricted rotation, we predict the form of the rotational correlation function using the method of Lipari and Szabo~\cite{lipari1981pade}.
This approach assumes that the molecule rotates freely within a cone of semiangle $\theta_0$, and then approximates the rotational time correlation function by a sum of weighted exponentials, where the weights involve the cone semiangle,
\begin{equation}
C_2(t)\approx \sum_{m=-2}^2 G_m(\infty) + F_m(\theta_0) e^{-t/\tau^{(m)}_{\rm eff}},
\end{equation}
where $\tau^{(m)}_{\rm eff}$ is an effective relaxation time.
The prefactors can be written in terms of Lipari and Szabo's $G_m(t;\cos\theta_0)$ functions as
\begin{equation}
F_m(\theta_0) = G_m(0;\cos\theta_0)-G_m(\infty;\cos\theta_0),
\end{equation}
where $G_m(\infty;\cos\theta_0) =  \delta_{m0} \brac{\frac{1}{2}\cos\theta_0 (1+\cos\theta_0)}^2$ and $\delta_{m0}$ is a Kronecker delta function.
The effective relaxation time is defined as $\tau_{\rm eff}^{(m)}=\tau_m / \brac{G_m(0;\cos\theta_0)-G_m(\infty;\cos\theta_0)}$.
The remaining $G_m(0;\cos\theta_0)$ and $\tau_m$ expressions follow exactly from Lipari and Szabo~\cite{lipari1981pade}, and we refer the reader there for the complete expressions.
The relaxation times, $\tau_m=\tau_m(\cos\theta_0,D_T)$, are functions of the semiangle and the rotational diffusion constant, $D_T$.
As a result, the only input to the theory from our simulations is an estimate of the rotational diffusion constant, $D_{T} \approx \frac{1}{l(l+1) \tau_T},$ where $T$ is the temperature, $l=2$ for $C_l(t)=C_2(t)$, and $\tau_T$ is equal to $\tau_{90}$ or $\tau_{30}$ at $T=90$~K or $T=30$~K, respectively, as determined from simulations.
We treat $\theta_0$ as a fit parameter that is adjusted so that $G_0(\infty;\cos\theta_0)$ closely matches $C_2(\infty)$.

The predicted $C_2(t)$ are shown as gray dashed lines in Fig.~\ref{fig:tcf} and the simulated data at 90~K agree well with the predictions for $\theta_0=63.5\degree$.
The good agreement of the prediction with the computed $C_2(t)$ suggests that the acetonitrile molecules in the crystal exhibit hindered but free rotation.
At 30~K, the $C_2(t)$ predicted by the Lipari and Szabo approach with the same $\theta_0=63.5\degree$ agrees less well with the simulation results, although the general timescale of the decay is closely approximated. 
Therefore, even at the low temperature of 30~K, the acetonitrile molecules within the co-crystal seem to undergo hindered but nearly free rotation.
We suggest that the presence and persistence of nearly free rotations over a wide range of temperatures is due to the lack of strong, specific interactions between the acetonitrile CH$_3$ group and other molecules in the co-crystal. 
%

%**********************************************************************************%
\subsection{Interaction Dynamics}

%------------------------------------------------------------------------------------------------------------------------------------------------------------------
\begin{figure*}[tb]
\begin{center}
\includegraphics[width=0.96\textwidth]{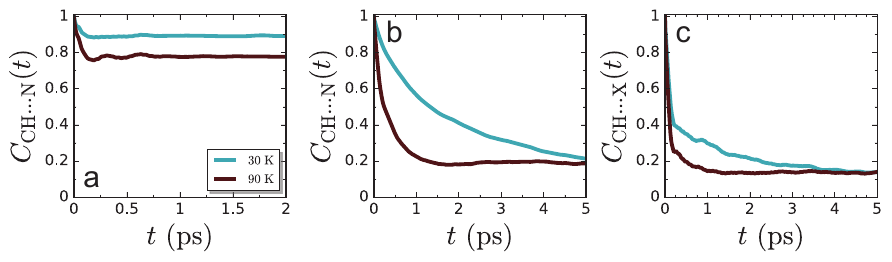}
\end{center}
\caption
{
Time correlation functions quantifying the lifetime of (a) acetylene C-H$\cdots$N hydrogen bonds,
(b) acetonitrile C-H$\cdots$N interactions, and (c) acetonitrile C-H$\cdots\pi$ interactions.
}
\label{fig:hb}
\end{figure*}
%------------------------------------------------------------------------------------------------------------------------------------------------------------------

%%
%
Rotation of the acetonitrile molecule arounds its $C_3$ symmetry axis must disrupt the weak, directional acetonitrile C-H$\cdots$N and C-H$\cdots\pi$ interactions, and thermal fluctuations in the solid can also disrupt acetylene C-H$\cdots$N hydrogen bonds.
As a result, these interactions should exhibit dynamics that mimic the molecular dynamics in the co-crystal.
We quantify the dynamics of these interactions through time correlation functions of the form
\begin{equation}
C(t) = \frac{\avg{h(0)h(t)}}{\avg{h}},
\end{equation}
where $h(t)=\Theta\brac{R_c-R(t)}\Theta\brac{\beta_c-\beta(t)}$ and $\Theta(x)$ is the Heaviside step function, such that
$h(t)=1$ if the geometric criterion for the interaction is satisfied at time $t$ and $h(t)=0$ otherwise.
The angle and distance cutoffs used to identify if an interaction is present at time $t$ are indicated by the dashed lines in the free energy surfaces in Fig.~\ref{fig:fes}.
For acetylene C-H$\cdots$N hydrogen bonds, we use $R_c=3.6$~\AA \ and $\beta_c=30\degree$, consistent with definitions used for hydrogen bonds in water~\cite{luzar1996hydrogen,kumar2007hydrogen}, for example.
For acetonitrile C-H$\cdots$N interactions, we use larger cutoffs of $R_c=3.8$~\AA \ and $\beta_c=50\degree$.
For acetonitrile C-H$\cdots\pi$ interactions, we use even larger cutoffs of $R_c=4$~\AA \ and $\beta_c=50\degree$.
The time correlation function quantifying acetylene C-H$\cdots$N hydrogen bonds only decays by about 10\% and 20\% at 30~K and 90~K, respectively, Fig.~\ref{fig:hb}a, indicating that these hydrogen bonds transiently break and reform.
The partial decay is due to thermal fluctuations, which are larger at 90~K, consistent with the merging of two minima in the free energy surface at this temperature, Fig.~\ref{fig:fes}a,d.
The time correlation function quantifying acetonitrile C-H$\cdots$N interactions, Fig.~\ref{fig:hb}b, decays to a long-time value of $\avg{h}\approx0.2$ at both 30~K and 90~K. 
The decays of the correlation functions are similar to that of the corresponding $C_2(t)$, Fig.~\ref{fig:tcf}b.
The similarity of the decay of interactions to the rotational decay is expected because the rotation of the acetonitrile CH$_3$ group must disrupt C-H$\cdots$N interactions. 
The time correlation function quantifying acetonitrile C-H$\cdots\pi$ interactions exhibits two timescales, Fig.~\ref{fig:hb}c.
The two timescales suggest two decay mechanisms. 
The fast decay at short times results from disruptions of the C-H$\cdots\pi$ interactions due to thermally excited phonons that increase the acetonitrile-acetylene distance, $R$.
The long time decay mirrors the decay of $C_2(t)$, indicating that the longest decay is a result of CH$_3$ rotations that disrupt C-H$\cdots\pi$ interactions by changing $\beta$. 
This increase in $\beta$ beyond the geometric cutoff defining the interaction occurs with a free energy change less than $\kT$, consistent with free rotation.
%

%**********************************************************************************%
\section{Conclusion}
We predict that the acetonitrile:acetylene co-crystal is a plastic crystal at Titan surface temperatures.
The acetonitrile molecules are orientationally disordered about their C-C axis ($C_3$ symmetry axis) and exhibit fast, sub-picosecond rotational dynamics. 
These fast rotational dynamics are consistent with free rotation confined within a cone, suggesting that weak directional interactions among acetonitrile and acetylene molecules are readily disrupted by thermal fluctuations and have little impact on dynamics within the co-crystal.
Our prediction of disorder across a wide temperature range is consistent with the presence of disorder in the experimental crystal structure~\cite{cable2020properties}.
Acetonitrile:acetylene is one member of a larger class of cryominerals predicted to exhibit disorder at Titan conditions~\cite{maynard2018prospects,thakur2023molecular,thakur2024nuclear}, which suggests that plastic crystals may be prevalent in the organic cryominerals on Titan's surface. 
The presence of disorder in organic crystals can significantly impact chemical reaction rates and even the products of chemical reactions~\cite{tran2005photochemical,gavezzotti1982crystal,shalaev2002effect}.
As a result, plastic crystal dynamics may lead to previously unexpected solid-state chemistry on Titan, further enriching its vast chemical inventory.
Plastic crystal disorder also alters the mechanical properties of materials, often through lowering specific elastic moduli along the directions of disorder~\cite{lynden1994translation,klein1990simulation}. 
Therefore, plastic crystal disorder may also play a key role in shaping the landscape of Titan's surface through weathering, fracture, and response to impact events~\cite{maynard2018prospects,NeishGRL2015,yu2018does,solomonidou2020chemical,yu2023material}. 
Building a quantitative understanding of the role of plastic crystals in these processes and more will be crucial to understanding the geochemistry of Titan and similar planetary bodies. 
%

%**********************************************************************************************************%
\begin{acknowledgements}
This work is supported by the National Aeronautics and Space Administration under grant number 80NSSC20K0609, issued through the NASA Exobiology Program.
We acknowledge the Office of Advanced Research Computing (OARC) at Rutgers,
The State University of New Jersey
for providing access to the Amarel cluster
and associated research computing resources that have contributed to the results reported here.
This work used the Advanced Cyberinfrastructure Coordination Ecosystem: Services \& Support (ACCESS), formerly Extreme Science and Engineering Discovery Environment (XSEDE)~\cite{towns2014xsede}, which is supported by National Science Foundation grant number ACI-1548562. Specifically, this work used Stampede2 and Ranch at the Texas Advanced Computing Center through allocation TG-CHE210081.
\end{acknowledgements}

\bibliography{AA}

\begin{thebibliography}{46}%
\makeatletter
\providecommand \@ifxundefined [1]{%
 \@ifx{#1\undefined}
}%
\providecommand \@ifnum [1]{%
 \ifnum #1\expandafter \@firstoftwo
 \else \expandafter \@secondoftwo
 \fi
}%
\providecommand \@ifx [1]{%
 \ifx #1\expandafter \@firstoftwo
 \else \expandafter \@secondoftwo
 \fi
}%
\providecommand \natexlab [1]{#1}%
\providecommand \enquote  [1]{``#1''}%
\providecommand \bibnamefont  [1]{#1}%
\providecommand \bibfnamefont [1]{#1}%
\providecommand \citenamefont [1]{#1}%
\providecommand \href@noop [0]{\@secondoftwo}%
\providecommand \href [0]{\begingroup \@sanitize@url \@href}%
\providecommand \@href[1]{\@@startlink{#1}\@@href}%
\providecommand \@@href[1]{\endgroup#1\@@endlink}%
\providecommand \@sanitize@url [0]{\catcode `\\12\catcode `\$12\catcode
  `\&12\catcode `\#12\catcode `\^12\catcode `\_12\catcode `\%12\relax}%
\providecommand \@@startlink[1]{}%
\providecommand \@@endlink[0]{}%
\providecommand \url  [0]{\begingroup\@sanitize@url \@url }%
\providecommand \@url [1]{\endgroup\@href {#1}{\urlprefix }}%
\providecommand \urlprefix  [0]{URL }%
\providecommand \Eprint [0]{\href }%
\providecommand \doibase [0]{http://dx.doi.org/}%
\providecommand \selectlanguage [0]{\@gobble}%
\providecommand \bibinfo  [0]{\@secondoftwo}%
\providecommand \bibfield  [0]{\@secondoftwo}%
\providecommand \translation [1]{[#1]}%
\providecommand \BibitemOpen [0]{}%
\providecommand \bibitemStop [0]{}%
\providecommand \bibitemNoStop [0]{.\EOS\space}%
\providecommand \EOS [0]{\spacefactor3000\relax}%
\providecommand \BibitemShut  [1]{\csname bibitem#1\endcsname}%
\let\auto@bib@innerbib\@empty
%</preamble>
\bibitem [{\citenamefont {Lorenz}\ \emph {et~al.}(2006)\citenamefont {Lorenz},
  \citenamefont {Wall}, \citenamefont {Radebaugh}, \citenamefont {Boubin},
  \citenamefont {Reffet}, \citenamefont {Janssen}, \citenamefont {Stofan},
  \citenamefont {Lopes}, \citenamefont {Kirk}, \citenamefont {Elachi},
  \citenamefont {Lunine}, \citenamefont {Mitchell}, \citenamefont {Paganelli},
  \citenamefont {Soderblom}, \citenamefont {Wood}, \citenamefont {Wye},
  \citenamefont {Zebker}, \citenamefont {Anderson}, \citenamefont {Ostro},
  \citenamefont {Allison}, \citenamefont {Boehmer}, \citenamefont {Callahan},
  \citenamefont {Encrenaz}, \citenamefont {Ori}, \citenamefont {Francescetti},
  \citenamefont {Gim}, \citenamefont {Hamilton}, \citenamefont {Hensley},
  \citenamefont {Johnson}, \citenamefont {Kelleher}, \citenamefont {Muhleman},
  \citenamefont {Picardi}, \citenamefont {Posa}, \citenamefont {Roth},
  \citenamefont {Seu}, \citenamefont {Shaffer}, \citenamefont {Stiles},
  \citenamefont {Vetrella}, \citenamefont {Flamini},\ and\ \citenamefont
  {West}}]{lorenz2006sand}%
  \BibitemOpen
  \bibfield  {author} {\bibinfo {author} {\bibfnamefont {R.~D.}\ \bibnamefont
  {Lorenz}}, \bibinfo {author} {\bibfnamefont {S.}~\bibnamefont {Wall}},
  \bibinfo {author} {\bibfnamefont {J.}~\bibnamefont {Radebaugh}}, \bibinfo
  {author} {\bibfnamefont {G.}~\bibnamefont {Boubin}}, \bibinfo {author}
  {\bibfnamefont {E.}~\bibnamefont {Reffet}}, \bibinfo {author} {\bibfnamefont
  {M.}~\bibnamefont {Janssen}}, \bibinfo {author} {\bibfnamefont
  {E.}~\bibnamefont {Stofan}}, \bibinfo {author} {\bibfnamefont
  {R.}~\bibnamefont {Lopes}}, \bibinfo {author} {\bibfnamefont
  {R.}~\bibnamefont {Kirk}}, \bibinfo {author} {\bibfnamefont {C.}~\bibnamefont
  {Elachi}}, \bibinfo {author} {\bibfnamefont {J.}~\bibnamefont {Lunine}},
  \bibinfo {author} {\bibfnamefont {K.}~\bibnamefont {Mitchell}}, \bibinfo
  {author} {\bibfnamefont {F.}~\bibnamefont {Paganelli}}, \bibinfo {author}
  {\bibfnamefont {L.}~\bibnamefont {Soderblom}}, \bibinfo {author}
  {\bibfnamefont {C.}~\bibnamefont {Wood}}, \bibinfo {author} {\bibfnamefont
  {L.}~\bibnamefont {Wye}}, \bibinfo {author} {\bibfnamefont {H.}~\bibnamefont
  {Zebker}}, \bibinfo {author} {\bibfnamefont {Y.}~\bibnamefont {Anderson}},
  \bibinfo {author} {\bibfnamefont {S.}~\bibnamefont {Ostro}}, \bibinfo
  {author} {\bibfnamefont {M.}~\bibnamefont {Allison}}, \bibinfo {author}
  {\bibfnamefont {R.}~\bibnamefont {Boehmer}}, \bibinfo {author} {\bibfnamefont
  {P.}~\bibnamefont {Callahan}}, \bibinfo {author} {\bibfnamefont
  {P.}~\bibnamefont {Encrenaz}}, \bibinfo {author} {\bibfnamefont {G.~G.}\
  \bibnamefont {Ori}}, \bibinfo {author} {\bibfnamefont {G.}~\bibnamefont
  {Francescetti}}, \bibinfo {author} {\bibfnamefont {Y.}~\bibnamefont {Gim}},
  \bibinfo {author} {\bibfnamefont {G.}~\bibnamefont {Hamilton}}, \bibinfo
  {author} {\bibfnamefont {S.}~\bibnamefont {Hensley}}, \bibinfo {author}
  {\bibfnamefont {W.}~\bibnamefont {Johnson}}, \bibinfo {author} {\bibfnamefont
  {K.}~\bibnamefont {Kelleher}}, \bibinfo {author} {\bibfnamefont
  {D.}~\bibnamefont {Muhleman}}, \bibinfo {author} {\bibfnamefont
  {G.}~\bibnamefont {Picardi}}, \bibinfo {author} {\bibfnamefont
  {F.}~\bibnamefont {Posa}}, \bibinfo {author} {\bibfnamefont {L.}~\bibnamefont
  {Roth}}, \bibinfo {author} {\bibfnamefont {R.}~\bibnamefont {Seu}}, \bibinfo
  {author} {\bibfnamefont {S.}~\bibnamefont {Shaffer}}, \bibinfo {author}
  {\bibfnamefont {B.}~\bibnamefont {Stiles}}, \bibinfo {author} {\bibfnamefont
  {S.}~\bibnamefont {Vetrella}}, \bibinfo {author} {\bibfnamefont
  {E.}~\bibnamefont {Flamini}}, \ and\ \bibinfo {author} {\bibfnamefont
  {R.}~\bibnamefont {West}},\ }\bibfield  {title} {\enquote {\bibinfo {title}
  {{T}he {S}and {S}eas of {T}itan: {C}assini {RADAR} {O}bservations of
  {L}ongitudinal {D}unes},}\ }\href@noop {} {\bibfield  {journal} {\bibinfo
  {journal} {Science}\ }\textbf {\bibinfo {volume} {312}},\ \bibinfo {pages}
  {724--727} (\bibinfo {year} {2006})}\BibitemShut {NoStop}%
\bibitem [{\citenamefont {Singh}\ \emph {et~al.}(2016)\citenamefont {Singh},
  \citenamefont {McCord}, \citenamefont {Combe}, \citenamefont {Rodriguez},
  \citenamefont {Cornet}, \citenamefont {Le~Mou{\'e}lic}, \citenamefont
  {Clark}, \citenamefont {Maltagliati},\ and\ \citenamefont
  {Chevrier}}]{singh2016acetylene}%
  \BibitemOpen
  \bibfield  {author} {\bibinfo {author} {\bibfnamefont {S.}~\bibnamefont
  {Singh}}, \bibinfo {author} {\bibfnamefont {T.}~\bibnamefont {McCord}},
  \bibinfo {author} {\bibfnamefont {J.-P.}\ \bibnamefont {Combe}}, \bibinfo
  {author} {\bibfnamefont {S.}~\bibnamefont {Rodriguez}}, \bibinfo {author}
  {\bibfnamefont {T.}~\bibnamefont {Cornet}}, \bibinfo {author} {\bibfnamefont
  {S.}~\bibnamefont {Le~Mou{\'e}lic}}, \bibinfo {author} {\bibfnamefont
  {R.}~\bibnamefont {Clark}}, \bibinfo {author} {\bibfnamefont
  {L.}~\bibnamefont {Maltagliati}}, \ and\ \bibinfo {author} {\bibfnamefont
  {V.}~\bibnamefont {Chevrier}},\ }\bibfield  {title} {\enquote {\bibinfo
  {title} {{A}cetylene on {T}itan's {S}urface},}\ }\href@noop {} {\bibfield
  {journal} {\bibinfo  {journal} {Astrophys. J.}\ }\textbf {\bibinfo {volume}
  {828}},\ \bibinfo {pages} {55} (\bibinfo {year} {2016})}\BibitemShut
  {NoStop}%
\bibitem [{\citenamefont {H{\"o}rst}(2017)}]{horst2017titan}%
  \BibitemOpen
  \bibfield  {author} {\bibinfo {author} {\bibfnamefont {S.~M.}\ \bibnamefont
  {H{\"o}rst}},\ }\bibfield  {title} {\enquote {\bibinfo {title} {{T}itan's
  {A}tmosphere and {C}limate},}\ }\href@noop {} {\bibfield  {journal} {\bibinfo
   {journal} {J. Geophys. Res.: Planets}\ }\textbf {\bibinfo {volume} {122}},\
  \bibinfo {pages} {432--482} (\bibinfo {year} {2017})}\BibitemShut {NoStop}%
\bibitem [{\citenamefont {Lunine}\ \emph {et~al.}(2020)\citenamefont {Lunine},
  \citenamefont {Cable}, \citenamefont {H{\"o}rst},\ and\ \citenamefont
  {Rahm}}]{lunine2020astrobiology}%
  \BibitemOpen
  \bibfield  {author} {\bibinfo {author} {\bibfnamefont {J.~I.}\ \bibnamefont
  {Lunine}}, \bibinfo {author} {\bibfnamefont {M.~L.}\ \bibnamefont {Cable}},
  \bibinfo {author} {\bibfnamefont {S.~M.}\ \bibnamefont {H{\"o}rst}}, \ and\
  \bibinfo {author} {\bibfnamefont {M.}~\bibnamefont {Rahm}},\ }\bibfield
  {title} {\enquote {\bibinfo {title} {{T}he {A}strobiology of {T}itan},}\
  }\href@noop {} {\bibfield  {journal} {\bibinfo  {journal} {Planetary
  Astrobiology}\ ,\ \bibinfo {pages} {247}} (\bibinfo {year}
  {2020})}\BibitemShut {NoStop}%
\bibitem [{\citenamefont {MacKenzie}\ \emph {et~al.}(2021)\citenamefont
  {MacKenzie}, \citenamefont {Birch}, \citenamefont {H{\"o}rst}, \citenamefont
  {Sotin}, \citenamefont {Barth}, \citenamefont {Lora}, \citenamefont
  {Trainer}, \citenamefont {Corlies}, \citenamefont {Malaska}, \citenamefont
  {Sciamma-O'Brien}, \citenamefont {Thelen}, \citenamefont {Turtle},
  \citenamefont {Radebaugh}, \citenamefont {Hanley}, \citenamefont
  {Solomonidou}, \citenamefont {Newman}, \citenamefont {Regoli}, \citenamefont
  {Rodriguez}, \citenamefont {Seignovert}, \citenamefont {Hayes}, \citenamefont
  {Journaux}, \citenamefont {Steckloff}, \citenamefont {Nna-Mvondo},
  \citenamefont {Cornet}, \citenamefont {Palmer}, \citenamefont {Lopes},
  \citenamefont {Vinatier}, \citenamefont {Lorenz}, \citenamefont {Nixon},
  \citenamefont {Czaplinski}, \citenamefont {Barnes}, \citenamefont {Sittler},\
  and\ \citenamefont {Coates}}]{mackenzie2021titan}%
  \BibitemOpen
  \bibfield  {author} {\bibinfo {author} {\bibfnamefont {S.~M.}\ \bibnamefont
  {MacKenzie}}, \bibinfo {author} {\bibfnamefont {S.~P.~D.}\ \bibnamefont
  {Birch}}, \bibinfo {author} {\bibfnamefont {S.}~\bibnamefont {H{\"o}rst}},
  \bibinfo {author} {\bibfnamefont {C.}~\bibnamefont {Sotin}}, \bibinfo
  {author} {\bibfnamefont {E.}~\bibnamefont {Barth}}, \bibinfo {author}
  {\bibfnamefont {J.~M.}\ \bibnamefont {Lora}}, \bibinfo {author}
  {\bibfnamefont {M.~G.}\ \bibnamefont {Trainer}}, \bibinfo {author}
  {\bibfnamefont {P.}~\bibnamefont {Corlies}}, \bibinfo {author} {\bibfnamefont
  {M.~J.}\ \bibnamefont {Malaska}}, \bibinfo {author} {\bibfnamefont
  {E.}~\bibnamefont {Sciamma-O'Brien}}, \bibinfo {author} {\bibfnamefont
  {A.~E.}\ \bibnamefont {Thelen}}, \bibinfo {author} {\bibfnamefont
  {E.}~\bibnamefont {Turtle}}, \bibinfo {author} {\bibfnamefont
  {J.}~\bibnamefont {Radebaugh}}, \bibinfo {author} {\bibfnamefont
  {J.}~\bibnamefont {Hanley}}, \bibinfo {author} {\bibfnamefont
  {A.}~\bibnamefont {Solomonidou}}, \bibinfo {author} {\bibfnamefont
  {C.}~\bibnamefont {Newman}}, \bibinfo {author} {\bibfnamefont
  {L.}~\bibnamefont {Regoli}}, \bibinfo {author} {\bibfnamefont
  {S.}~\bibnamefont {Rodriguez}}, \bibinfo {author} {\bibfnamefont
  {B.}~\bibnamefont {Seignovert}}, \bibinfo {author} {\bibfnamefont {A.~G.}\
  \bibnamefont {Hayes}}, \bibinfo {author} {\bibfnamefont {B.}~\bibnamefont
  {Journaux}}, \bibinfo {author} {\bibfnamefont {J.}~\bibnamefont {Steckloff}},
  \bibinfo {author} {\bibfnamefont {D.}~\bibnamefont {Nna-Mvondo}}, \bibinfo
  {author} {\bibfnamefont {T.}~\bibnamefont {Cornet}}, \bibinfo {author}
  {\bibfnamefont {M.~Y.}\ \bibnamefont {Palmer}}, \bibinfo {author}
  {\bibfnamefont {R.~M.~C.}\ \bibnamefont {Lopes}}, \bibinfo {author}
  {\bibfnamefont {S.}~\bibnamefont {Vinatier}}, \bibinfo {author}
  {\bibfnamefont {R.}~\bibnamefont {Lorenz}}, \bibinfo {author} {\bibfnamefont
  {C.}~\bibnamefont {Nixon}}, \bibinfo {author} {\bibfnamefont
  {E.}~\bibnamefont {Czaplinski}}, \bibinfo {author} {\bibfnamefont {J.~W.}\
  \bibnamefont {Barnes}}, \bibinfo {author} {\bibfnamefont {E.}~\bibnamefont
  {Sittler}}, \ and\ \bibinfo {author} {\bibfnamefont {A.}~\bibnamefont
  {Coates}},\ }\bibfield  {title} {\enquote {\bibinfo {title} {{T}itan:
  {E}arth-like on the {O}utside, {O}cean {W}orld on the {I}nside},}\
  }\href@noop {} {\bibfield  {journal} {\bibinfo  {journal} {Planet. Sci. J.}\
  }\textbf {\bibinfo {volume} {2}},\ \bibinfo {pages} {112} (\bibinfo {year}
  {2021})}\BibitemShut {NoStop}%
\bibitem [{\citenamefont {Nixon}(2024)}]{nixon2024composition}%
  \BibitemOpen
  \bibfield  {author} {\bibinfo {author} {\bibfnamefont {C.~A.}\ \bibnamefont
  {Nixon}},\ }\bibfield  {title} {\enquote {\bibinfo {title} {The composition
  and chemistry of titan's atmosphere},}\ }\href@noop {} {\bibfield  {journal}
  {\bibinfo  {journal} {ACS Earth and Space Chemistry}\ }\textbf {\bibinfo
  {volume} {8}},\ \bibinfo {pages} {406--456} (\bibinfo {year}
  {2024})}\BibitemShut {NoStop}%
\bibitem [{\citenamefont {Maynard-Casely}\ \emph {et~al.}(2018)\citenamefont
  {Maynard-Casely}, \citenamefont {Cable}, \citenamefont {Malaska},
  \citenamefont {Vu}, \citenamefont {Choukroun},\ and\ \citenamefont
  {Hodyss}}]{maynard2018prospects}%
  \BibitemOpen
  \bibfield  {author} {\bibinfo {author} {\bibfnamefont {H.~E.}\ \bibnamefont
  {Maynard-Casely}}, \bibinfo {author} {\bibfnamefont {M.~L.}\ \bibnamefont
  {Cable}}, \bibinfo {author} {\bibfnamefont {M.~J.}\ \bibnamefont {Malaska}},
  \bibinfo {author} {\bibfnamefont {T.~H.}\ \bibnamefont {Vu}}, \bibinfo
  {author} {\bibfnamefont {M.}~\bibnamefont {Choukroun}}, \ and\ \bibinfo
  {author} {\bibfnamefont {R.}~\bibnamefont {Hodyss}},\ }\bibfield  {title}
  {\enquote {\bibinfo {title} {{P}rospects for {M}ineralogy on {T}itan},}\
  }\href@noop {} {\bibfield  {journal} {\bibinfo  {journal} {Am. Mineral.}\
  }\textbf {\bibinfo {volume} {103}},\ \bibinfo {pages} {343--349} (\bibinfo
  {year} {2018})}\BibitemShut {NoStop}%
\bibitem [{\citenamefont {Cable}\ \emph {et~al.}(2021)\citenamefont {Cable},
  \citenamefont {Run{\v c}evski}, \citenamefont {Maynard-Casely}, \citenamefont
  {Vu},\ and\ \citenamefont {Hodyss}}]{cable2021titan}%
  \BibitemOpen
  \bibfield  {author} {\bibinfo {author} {\bibfnamefont {M.~L.}\ \bibnamefont
  {Cable}}, \bibinfo {author} {\bibfnamefont {T.}~\bibnamefont {Run{\v
  c}evski}}, \bibinfo {author} {\bibfnamefont {H.~E.}\ \bibnamefont
  {Maynard-Casely}}, \bibinfo {author} {\bibfnamefont {T.~H.}\ \bibnamefont
  {Vu}}, \ and\ \bibinfo {author} {\bibfnamefont {R.}~\bibnamefont {Hodyss}},\
  }\bibfield  {title} {\enquote {\bibinfo {title} {{T}itan in a {T}est {T}ube:
  {O}rganic {C}o-crystals and {I}mplications for {T}itan {M}ineralogy},}\
  }\href@noop {} {\bibfield  {journal} {\bibinfo  {journal} {Acc. Chem. Res.}\
  }\textbf {\bibinfo {volume} {54}},\ \bibinfo {pages} {3050--3059} (\bibinfo
  {year} {2021})}\BibitemShut {NoStop}%
\bibitem [{\citenamefont {McConville}\ \emph {et~al.}(2020)\citenamefont
  {McConville}, \citenamefont {Tao}, \citenamefont {Evans}, \citenamefont
  {Trump}, \citenamefont {Lefton}, \citenamefont {Xu}, \citenamefont
  {Yakovenko}, \citenamefont {Kraka}, \citenamefont {Brown},\ and\
  \citenamefont {Run{\v{c}}evski}}]{mcconville2020peritectic}%
  \BibitemOpen
  \bibfield  {author} {\bibinfo {author} {\bibfnamefont {C.~A.}\ \bibnamefont
  {McConville}}, \bibinfo {author} {\bibfnamefont {Y.}~\bibnamefont {Tao}},
  \bibinfo {author} {\bibfnamefont {H.~A.}\ \bibnamefont {Evans}}, \bibinfo
  {author} {\bibfnamefont {B.~A.}\ \bibnamefont {Trump}}, \bibinfo {author}
  {\bibfnamefont {J.~B.}\ \bibnamefont {Lefton}}, \bibinfo {author}
  {\bibfnamefont {W.}~\bibnamefont {Xu}}, \bibinfo {author} {\bibfnamefont
  {A.~A.}\ \bibnamefont {Yakovenko}}, \bibinfo {author} {\bibfnamefont
  {E.}~\bibnamefont {Kraka}}, \bibinfo {author} {\bibfnamefont {C.~M.}\
  \bibnamefont {Brown}}, \ and\ \bibinfo {author} {\bibfnamefont
  {T.}~\bibnamefont {Run{\v{c}}evski}},\ }\bibfield  {title} {\enquote
  {\bibinfo {title} {Peritectic phase transition of benzene and acetonitrile
  into a cocrystal relevant to titan, saturn's moon},}\ }\href@noop {}
  {\bibfield  {journal} {\bibinfo  {journal} {Chemical Communications}\
  }\textbf {\bibinfo {volume} {56}},\ \bibinfo {pages} {13520--13523} (\bibinfo
  {year} {2020})}\BibitemShut {NoStop}%
\bibitem [{\citenamefont {Klein}\ and\ \citenamefont
  {Lewis}(1990)}]{klein1990simulation}%
  \BibitemOpen
  \bibfield  {author} {\bibinfo {author} {\bibfnamefont {M.~L.}\ \bibnamefont
  {Klein}}\ and\ \bibinfo {author} {\bibfnamefont {L.~J.}\ \bibnamefont
  {Lewis}},\ }\bibfield  {title} {\enquote {\bibinfo {title} {{S}imulation of
  {D}ynamical {P}rocesses in {M}olecular {S}olids},}\ }\href@noop {} {\bibfield
   {journal} {\bibinfo  {journal} {Chem. Rev.}\ }\textbf {\bibinfo {volume}
  {90}},\ \bibinfo {pages} {459--479} (\bibinfo {year} {1990})}\BibitemShut
  {NoStop}%
\bibitem [{\citenamefont {Klein}(1985)}]{klein1985computer}%
  \BibitemOpen
  \bibfield  {author} {\bibinfo {author} {\bibfnamefont {M.~L.}\ \bibnamefont
  {Klein}},\ }\bibfield  {title} {\enquote {\bibinfo {title} {Computer
  simulation studies of solids},}\ }\href@noop {} {\bibfield  {journal}
  {\bibinfo  {journal} {Annual Review of Physical Chemistry}\ }\textbf
  {\bibinfo {volume} {36}},\ \bibinfo {pages} {525--548} (\bibinfo {year}
  {1985})}\BibitemShut {NoStop}%
\bibitem [{\citenamefont {Lynden-Bell}\ and\ \citenamefont
  {Michel}(1994)}]{lynden1994translation}%
  \BibitemOpen
  \bibfield  {author} {\bibinfo {author} {\bibfnamefont {R.}~\bibnamefont
  {Lynden-Bell}}\ and\ \bibinfo {author} {\bibfnamefont {K.}~\bibnamefont
  {Michel}},\ }\bibfield  {title} {\enquote {\bibinfo {title}
  {{T}ranslation-rotation {C}oupling, {P}hase {T}ransitions, and {E}lastic
  {P}henomena in {O}rientationally {D}isordered {C}rystals},}\ }\href@noop {}
  {\bibfield  {journal} {\bibinfo  {journal} {Rev. Modern Phys.}\ }\textbf
  {\bibinfo {volume} {66}},\ \bibinfo {pages} {721} (\bibinfo {year}
  {1994})}\BibitemShut {NoStop}%
\bibitem [{\citenamefont {Tran}\ \emph {et~al.}(2005)\citenamefont {Tran},
  \citenamefont {Joseph}, \citenamefont {Force}, \citenamefont {Briggs},
  \citenamefont {Vuitton},\ and\ \citenamefont
  {Ferris}}]{tran2005photochemical}%
  \BibitemOpen
  \bibfield  {author} {\bibinfo {author} {\bibfnamefont {B.~N.}\ \bibnamefont
  {Tran}}, \bibinfo {author} {\bibfnamefont {J.~C.}\ \bibnamefont {Joseph}},
  \bibinfo {author} {\bibfnamefont {M.}~\bibnamefont {Force}}, \bibinfo
  {author} {\bibfnamefont {R.~G.}\ \bibnamefont {Briggs}}, \bibinfo {author}
  {\bibfnamefont {V.}~\bibnamefont {Vuitton}}, \ and\ \bibinfo {author}
  {\bibfnamefont {J.~P.}\ \bibnamefont {Ferris}},\ }\bibfield  {title}
  {\enquote {\bibinfo {title} {Photochemical processes on titan: Irradiation of
  mixtures of gases that simulate titan's atmosphere},}\ }\href@noop {}
  {\bibfield  {journal} {\bibinfo  {journal} {Icarus}\ }\textbf {\bibinfo
  {volume} {177}},\ \bibinfo {pages} {106--115} (\bibinfo {year}
  {2005})}\BibitemShut {NoStop}%
\bibitem [{\citenamefont {Gavezzotti}\ and\ \citenamefont
  {Simonetta}(1982)}]{gavezzotti1982crystal}%
  \BibitemOpen
  \bibfield  {author} {\bibinfo {author} {\bibfnamefont {A.}~\bibnamefont
  {Gavezzotti}}\ and\ \bibinfo {author} {\bibfnamefont {M.}~\bibnamefont
  {Simonetta}},\ }\bibfield  {title} {\enquote {\bibinfo {title} {Crystal
  chemistry in organic solids},}\ }\href@noop {} {\bibfield  {journal}
  {\bibinfo  {journal} {Chemical Reviews}\ }\textbf {\bibinfo {volume} {82}},\
  \bibinfo {pages} {1--13} (\bibinfo {year} {1982})}\BibitemShut {NoStop}%
\bibitem [{\citenamefont {Shalaev}, \citenamefont {Shalaeva},\ and\
  \citenamefont {Zografi}(2002)}]{shalaev2002effect}%
  \BibitemOpen
  \bibfield  {author} {\bibinfo {author} {\bibfnamefont {E.}~\bibnamefont
  {Shalaev}}, \bibinfo {author} {\bibfnamefont {M.}~\bibnamefont {Shalaeva}}, \
  and\ \bibinfo {author} {\bibfnamefont {G.}~\bibnamefont {Zografi}},\
  }\bibfield  {title} {\enquote {\bibinfo {title} {The effect of disorder on
  the chemical reactivity of an organic solid, tetraglycine methyl ester:
  Change of the reaction mechanism},}\ }\href@noop {} {\bibfield  {journal}
  {\bibinfo  {journal} {Journal of pharmaceutical sciences}\ }\textbf {\bibinfo
  {volume} {91}},\ \bibinfo {pages} {584--593} (\bibinfo {year}
  {2002})}\BibitemShut {NoStop}%
\bibitem [{\citenamefont {Zhang}\ and\ \citenamefont
  {Nazar}(2022)}]{zhang2022exploiting}%
  \BibitemOpen
  \bibfield  {author} {\bibinfo {author} {\bibfnamefont {Z.}~\bibnamefont
  {Zhang}}\ and\ \bibinfo {author} {\bibfnamefont {L.~F.}\ \bibnamefont
  {Nazar}},\ }\bibfield  {title} {\enquote {\bibinfo {title} {Exploiting the
  paddle-wheel mechanism for the design of fast ion conductors},}\ }\href@noop
  {} {\bibfield  {journal} {\bibinfo  {journal} {Nature Reviews Materials}\
  }\textbf {\bibinfo {volume} {7}},\ \bibinfo {pages} {389--405} (\bibinfo
  {year} {2022})}\BibitemShut {NoStop}%
\bibitem [{\citenamefont {Dhattarwal}, \citenamefont {Somni},\ and\
  \citenamefont {Remsing}(2024)}]{dhattarwal2024electronic}%
  \BibitemOpen
  \bibfield  {author} {\bibinfo {author} {\bibfnamefont {H.~S.}\ \bibnamefont
  {Dhattarwal}}, \bibinfo {author} {\bibfnamefont {R.}~\bibnamefont {Somni}}, \
  and\ \bibinfo {author} {\bibfnamefont {R.~C.}\ \bibnamefont {Remsing}},\
  }\bibfield  {title} {\enquote {\bibinfo {title} {Electronic paddle-wheels in
  a solid-state electrolyte},}\ }\href@noop {} {\bibfield  {journal} {\bibinfo
  {journal} {Nature communications}\ }\textbf {\bibinfo {volume} {15}},\
  \bibinfo {pages} {121} (\bibinfo {year} {2024})}\BibitemShut {NoStop}%
\bibitem [{\citenamefont {Thakur}\ and\ \citenamefont
  {Remsing}(2023)}]{thakur2023molecular}%
  \BibitemOpen
  \bibfield  {author} {\bibinfo {author} {\bibfnamefont {A.~C.}\ \bibnamefont
  {Thakur}}\ and\ \bibinfo {author} {\bibfnamefont {R.~C.}\ \bibnamefont
  {Remsing}},\ }\bibfield  {title} {\enquote {\bibinfo {title} {Molecular
  structure, dynamics, and vibrational spectroscopy of the acetylene: Ammonia
  (1: 1) plastic co-crystal at titan conditions},}\ }\href@noop {} {\bibfield
  {journal} {\bibinfo  {journal} {ACS Earth and Space Chemistry}\ }\textbf
  {\bibinfo {volume} {7}},\ \bibinfo {pages} {479--489} (\bibinfo {year}
  {2023})}\BibitemShut {NoStop}%
\bibitem [{\citenamefont {Thakur}\ and\ \citenamefont
  {Remsing}(2024)}]{thakur2024nuclear}%
  \BibitemOpen
  \bibfield  {author} {\bibinfo {author} {\bibfnamefont {A.~C.}\ \bibnamefont
  {Thakur}}\ and\ \bibinfo {author} {\bibfnamefont {R.~C.}\ \bibnamefont
  {Remsing}},\ }\bibfield  {title} {\enquote {\bibinfo {title} {Nuclear quantum
  effects in the acetylene: ammonia plastic co-crystal},}\ }\href@noop {}
  {\bibfield  {journal} {\bibinfo  {journal} {The Journal of Chemical Physics}\
  }\textbf {\bibinfo {volume} {160}} (\bibinfo {year} {2024})}\BibitemShut
  {NoStop}%
\bibitem [{\citenamefont {Cable}\ \emph {et~al.}(2020)\citenamefont {Cable},
  \citenamefont {Vu}, \citenamefont {Malaska}, \citenamefont {Maynard-Casely},
  \citenamefont {Choukroun},\ and\ \citenamefont
  {Hodyss}}]{cable2020properties}%
  \BibitemOpen
  \bibfield  {author} {\bibinfo {author} {\bibfnamefont {M.~L.}\ \bibnamefont
  {Cable}}, \bibinfo {author} {\bibfnamefont {T.~H.}\ \bibnamefont {Vu}},
  \bibinfo {author} {\bibfnamefont {M.~J.}\ \bibnamefont {Malaska}}, \bibinfo
  {author} {\bibfnamefont {H.~E.}\ \bibnamefont {Maynard-Casely}}, \bibinfo
  {author} {\bibfnamefont {M.}~\bibnamefont {Choukroun}}, \ and\ \bibinfo
  {author} {\bibfnamefont {R.}~\bibnamefont {Hodyss}},\ }\bibfield  {title}
  {\enquote {\bibinfo {title} {{P}roperties and {B}ehavior of the
  {A}cetonitrile-{A}cetylene {C}o-crystal under {T}itan {S}urface
  {C}onditions},}\ }\href@noop {} {\bibfield  {journal} {\bibinfo  {journal}
  {ACS Earth Space Chem.}\ }\textbf {\bibinfo {volume} {4}},\ \bibinfo {pages}
  {1375--1385} (\bibinfo {year} {2020})}\BibitemShut {NoStop}%
\bibitem [{\citenamefont {VandeVondele}\ and\ \citenamefont
  {Hutter}(2003)}]{vandevondele2003efficient}%
  \BibitemOpen
  \bibfield  {author} {\bibinfo {author} {\bibfnamefont {J.}~\bibnamefont
  {VandeVondele}}\ and\ \bibinfo {author} {\bibfnamefont {J.}~\bibnamefont
  {Hutter}},\ }\bibfield  {title} {\enquote {\bibinfo {title} {{A}n {E}fficient
  {O}rbital {T}ransformation {M}ethod for {E}lectronic {S}tructure
  {C}alculations},}\ }\href@noop {} {\bibfield  {journal} {\bibinfo  {journal}
  {J. Chem. Phys.}\ }\textbf {\bibinfo {volume} {118}},\ \bibinfo {pages}
  {4365--4369} (\bibinfo {year} {2003})}\BibitemShut {NoStop}%
\bibitem [{\citenamefont {Hutter}\ \emph {et~al.}(2014)\citenamefont {Hutter},
  \citenamefont {Iannuzzi}, \citenamefont {Schiffmann},\ and\ \citenamefont
  {VandeVondele}}]{hutter2014cp2k}%
  \BibitemOpen
  \bibfield  {author} {\bibinfo {author} {\bibfnamefont {J.}~\bibnamefont
  {Hutter}}, \bibinfo {author} {\bibfnamefont {M.}~\bibnamefont {Iannuzzi}},
  \bibinfo {author} {\bibfnamefont {F.}~\bibnamefont {Schiffmann}}, \ and\
  \bibinfo {author} {\bibfnamefont {J.}~\bibnamefont {VandeVondele}},\
  }\bibfield  {title} {\enquote {\bibinfo {title} {{CP2K}: {A}tomistic
  {S}imulations of {C}ondensed {M}atter {S}ystems},}\ }\href@noop {} {\bibfield
   {journal} {\bibinfo  {journal} {WIREs Comput. Mol. Sci.}\ }\textbf {\bibinfo
  {volume} {4}},\ \bibinfo {pages} {15--25} (\bibinfo {year}
  {2014})}\BibitemShut {NoStop}%
\bibitem [{\citenamefont {K{\"u}hne}\ \emph {et~al.}(2020)\citenamefont
  {K{\"u}hne}, \citenamefont {Iannuzzi}, \citenamefont {Del~Ben}, \citenamefont
  {Rybkin}, \citenamefont {Seewald}, \citenamefont {Stein}, \citenamefont
  {Laino}, \citenamefont {Khaliullin}, \citenamefont {Sch{\"u}tt},
  \citenamefont {Schiffmann}, \citenamefont {Golze}, \citenamefont {Wilhelm},
  \citenamefont {Chulkov}, \citenamefont {Bani-Hashemian}, \citenamefont
  {Weber}, \citenamefont {Bor{\v s}tnik}, \citenamefont {Taillefumier},
  \citenamefont {Jakobovits}, \citenamefont {Lazzaro}, \citenamefont {Pabst},
  \citenamefont {M{\"u}ller}, \citenamefont {Schade}, \citenamefont {Guidon},
  \citenamefont {Andermatt}, \citenamefont {Holmberg}, \citenamefont
  {Schenter}, \citenamefont {Hehn}, \citenamefont {Bussy}, \citenamefont
  {Belleflamme}, \citenamefont {Tabacchi}, \citenamefont {Gl{\"o}{\ss}},
  \citenamefont {Lass}, \citenamefont {Bethune}, \citenamefont {Mundy},
  \citenamefont {Plessl}, \citenamefont {Watkins}, \citenamefont
  {VandeVondele}, \citenamefont {Krack},\ and\ \citenamefont {Hutter}}]{CP2K}%
  \BibitemOpen
  \bibfield  {author} {\bibinfo {author} {\bibfnamefont {T.~D.}\ \bibnamefont
  {K{\"u}hne}}, \bibinfo {author} {\bibfnamefont {M.}~\bibnamefont {Iannuzzi}},
  \bibinfo {author} {\bibfnamefont {M.}~\bibnamefont {Del~Ben}}, \bibinfo
  {author} {\bibfnamefont {V.~V.}\ \bibnamefont {Rybkin}}, \bibinfo {author}
  {\bibfnamefont {P.}~\bibnamefont {Seewald}}, \bibinfo {author} {\bibfnamefont
  {F.}~\bibnamefont {Stein}}, \bibinfo {author} {\bibfnamefont
  {T.}~\bibnamefont {Laino}}, \bibinfo {author} {\bibfnamefont {R.~Z.}\
  \bibnamefont {Khaliullin}}, \bibinfo {author} {\bibfnamefont
  {O.}~\bibnamefont {Sch{\"u}tt}}, \bibinfo {author} {\bibfnamefont
  {F.}~\bibnamefont {Schiffmann}}, \bibinfo {author} {\bibfnamefont
  {D.}~\bibnamefont {Golze}}, \bibinfo {author} {\bibfnamefont
  {J.}~\bibnamefont {Wilhelm}}, \bibinfo {author} {\bibfnamefont
  {S.}~\bibnamefont {Chulkov}}, \bibinfo {author} {\bibfnamefont {M.~H.}\
  \bibnamefont {Bani-Hashemian}}, \bibinfo {author} {\bibfnamefont
  {V.}~\bibnamefont {Weber}}, \bibinfo {author} {\bibfnamefont
  {U.}~\bibnamefont {Bor{\v s}tnik}}, \bibinfo {author} {\bibfnamefont
  {M.}~\bibnamefont {Taillefumier}}, \bibinfo {author} {\bibfnamefont {A.~S.}\
  \bibnamefont {Jakobovits}}, \bibinfo {author} {\bibfnamefont
  {A.}~\bibnamefont {Lazzaro}}, \bibinfo {author} {\bibfnamefont
  {H.}~\bibnamefont {Pabst}}, \bibinfo {author} {\bibfnamefont
  {T.}~\bibnamefont {M{\"u}ller}}, \bibinfo {author} {\bibfnamefont
  {R.}~\bibnamefont {Schade}}, \bibinfo {author} {\bibfnamefont
  {M.}~\bibnamefont {Guidon}}, \bibinfo {author} {\bibfnamefont
  {S.}~\bibnamefont {Andermatt}}, \bibinfo {author} {\bibfnamefont
  {N.}~\bibnamefont {Holmberg}}, \bibinfo {author} {\bibfnamefont {G.~K.}\
  \bibnamefont {Schenter}}, \bibinfo {author} {\bibfnamefont {A.}~\bibnamefont
  {Hehn}}, \bibinfo {author} {\bibfnamefont {A.}~\bibnamefont {Bussy}},
  \bibinfo {author} {\bibfnamefont {F.}~\bibnamefont {Belleflamme}}, \bibinfo
  {author} {\bibfnamefont {G.}~\bibnamefont {Tabacchi}}, \bibinfo {author}
  {\bibfnamefont {A.}~\bibnamefont {Gl{\"o}{\ss}}}, \bibinfo {author}
  {\bibfnamefont {M.}~\bibnamefont {Lass}}, \bibinfo {author} {\bibfnamefont
  {I.}~\bibnamefont {Bethune}}, \bibinfo {author} {\bibfnamefont {C.~J.}\
  \bibnamefont {Mundy}}, \bibinfo {author} {\bibfnamefont {C.}~\bibnamefont
  {Plessl}}, \bibinfo {author} {\bibfnamefont {M.}~\bibnamefont {Watkins}},
  \bibinfo {author} {\bibfnamefont {J.}~\bibnamefont {VandeVondele}}, \bibinfo
  {author} {\bibfnamefont {M.}~\bibnamefont {Krack}}, \ and\ \bibinfo {author}
  {\bibfnamefont {J.}~\bibnamefont {Hutter}},\ }\bibfield  {title} {\enquote
  {\bibinfo {title} {{CP2K}: {A}n {E}lectronic {S}tructure and {M}olecular
  {D}ynamics {S}oftware {P}ackage-{Q}uickstep: {E}fficient and {A}ccurate
  {E}lectronic {S}tructure {C}alculations},}\ }\href@noop {} {\bibfield
  {journal} {\bibinfo  {journal} {J. Chem. Phys.}\ }\textbf {\bibinfo {volume}
  {152}},\ \bibinfo {pages} {194103} (\bibinfo {year} {2020})}\BibitemShut
  {NoStop}%
\bibitem [{\citenamefont {Lippert}, \citenamefont {Hutter},\ and\ \citenamefont
  {Parrinello}(1997)}]{lippert1997hybrid}%
  \BibitemOpen
  \bibfield  {author} {\bibinfo {author} {\bibfnamefont {G.}~\bibnamefont
  {Lippert}}, \bibinfo {author} {\bibfnamefont {J.}~\bibnamefont {Hutter}}, \
  and\ \bibinfo {author} {\bibfnamefont {M.}~\bibnamefont {Parrinello}},\
  }\bibfield  {title} {\enquote {\bibinfo {title} {{A} {H}ybrid {G}aussian and
  {P}lane {W}ave {D}ensity {F}unctional {S}cheme},}\ }\href@noop {} {\bibfield
  {journal} {\bibinfo  {journal} {Mol. Phys.}\ }\textbf {\bibinfo {volume}
  {92}},\ \bibinfo {pages} {477--488} (\bibinfo {year} {1997})}\BibitemShut
  {NoStop}%
\bibitem [{\citenamefont {VandeVondele}\ and\ \citenamefont
  {Hutter}(2007)}]{vandevondele2007gaussian}%
  \BibitemOpen
  \bibfield  {author} {\bibinfo {author} {\bibfnamefont {J.}~\bibnamefont
  {VandeVondele}}\ and\ \bibinfo {author} {\bibfnamefont {J.}~\bibnamefont
  {Hutter}},\ }\bibfield  {title} {\enquote {\bibinfo {title} {{G}aussian
  {B}asis {S}ets for {A}ccurate {C}alculations on {M}olecular {S}ystems in
  {G}as and {C}ondensed {P}hases},}\ }\href@noop {} {\bibfield  {journal}
  {\bibinfo  {journal} {J. Chem. Phys.}\ }\textbf {\bibinfo {volume} {127}},\
  \bibinfo {pages} {114105} (\bibinfo {year} {2007})}\BibitemShut {NoStop}%
\bibitem [{\citenamefont {Hartwigsen}, \citenamefont {G{\oe}decker},\ and\
  \citenamefont {Hutter}(1998)}]{hartwigsen1998relativistic}%
  \BibitemOpen
  \bibfield  {author} {\bibinfo {author} {\bibfnamefont {C.}~\bibnamefont
  {Hartwigsen}}, \bibinfo {author} {\bibfnamefont {S.}~\bibnamefont
  {G{\oe}decker}}, \ and\ \bibinfo {author} {\bibfnamefont {J.}~\bibnamefont
  {Hutter}},\ }\bibfield  {title} {\enquote {\bibinfo {title} {{R}elativistic
  {S}eparable {D}ual-space {G}aussian {P}seudopotentials from {H} to {Rn}},}\
  }\href@noop {} {\bibfield  {journal} {\bibinfo  {journal} {Phys. Rev. B}\
  }\textbf {\bibinfo {volume} {58}},\ \bibinfo {pages} {3641} (\bibinfo {year}
  {1998})}\BibitemShut {NoStop}%
\bibitem [{\citenamefont {Goedecker}, \citenamefont {Teter},\ and\
  \citenamefont {Hutter}(1996)}]{goedecker1996separable}%
  \BibitemOpen
  \bibfield  {author} {\bibinfo {author} {\bibfnamefont {S.}~\bibnamefont
  {Goedecker}}, \bibinfo {author} {\bibfnamefont {M.}~\bibnamefont {Teter}}, \
  and\ \bibinfo {author} {\bibfnamefont {J.}~\bibnamefont {Hutter}},\
  }\bibfield  {title} {\enquote {\bibinfo {title} {{S}eparable {D}ual-space
  {G}aussian {P}seudopotentials},}\ }\href@noop {} {\bibfield  {journal}
  {\bibinfo  {journal} {Phys. Rev. B}\ }\textbf {\bibinfo {volume} {54}},\
  \bibinfo {pages} {1703} (\bibinfo {year} {1996})}\BibitemShut {NoStop}%
\bibitem [{\citenamefont {Krack}(2005)}]{krack2005pseudopotentials}%
  \BibitemOpen
  \bibfield  {author} {\bibinfo {author} {\bibfnamefont {M.}~\bibnamefont
  {Krack}},\ }\bibfield  {title} {\enquote {\bibinfo {title}
  {{P}seudopotentials for {H} to {Kr} {O}ptimized for {G}radient-corrected
  {E}xchange-correlation {F}unctionals},}\ }\href@noop {} {\bibfield  {journal}
  {\bibinfo  {journal} {Theor. Chem. Acc.}\ }\textbf {\bibinfo {volume}
  {114}},\ \bibinfo {pages} {145--152} (\bibinfo {year} {2005})}\BibitemShut
  {NoStop}%
\bibitem [{\citenamefont {Perdew}, \citenamefont {Burke},\ and\ \citenamefont
  {Ernzerhof}(1996)}]{perdew1996generalized}%
  \BibitemOpen
  \bibfield  {author} {\bibinfo {author} {\bibfnamefont {J.~P.}\ \bibnamefont
  {Perdew}}, \bibinfo {author} {\bibfnamefont {K.}~\bibnamefont {Burke}}, \
  and\ \bibinfo {author} {\bibfnamefont {M.}~\bibnamefont {Ernzerhof}},\
  }\bibfield  {title} {\enquote {\bibinfo {title} {{G}eneralized {G}radient
  {A}pproximation {M}ade {S}imple},}\ }\href@noop {} {\bibfield  {journal}
  {\bibinfo  {journal} {Phys. Rev. Lett.}\ }\textbf {\bibinfo {volume} {77}},\
  \bibinfo {pages} {3865} (\bibinfo {year} {1996})}\BibitemShut {NoStop}%
\bibitem [{\citenamefont {Grimme}\ \emph {et~al.}(2010)\citenamefont {Grimme},
  \citenamefont {Antony}, \citenamefont {Ehrlich},\ and\ \citenamefont
  {Krieg}}]{grimme2010consistent}%
  \BibitemOpen
  \bibfield  {author} {\bibinfo {author} {\bibfnamefont {S.}~\bibnamefont
  {Grimme}}, \bibinfo {author} {\bibfnamefont {J.}~\bibnamefont {Antony}},
  \bibinfo {author} {\bibfnamefont {S.}~\bibnamefont {Ehrlich}}, \ and\
  \bibinfo {author} {\bibfnamefont {H.}~\bibnamefont {Krieg}},\ }\bibfield
  {title} {\enquote {\bibinfo {title} {{A} {C}onsistent and {A}ccurate {A}b
  {I}nitio {P}arametrization of {D}ensity {F}unctional {D}ispersion
  {C}orrection {(DFT-D)} for the 94 {E}lements {H-Pu}},}\ }\href@noop {}
  {\bibfield  {journal} {\bibinfo  {journal} {J. Chem. Phys.}\ }\textbf
  {\bibinfo {volume} {132}},\ \bibinfo {pages} {154104} (\bibinfo {year}
  {2010})}\BibitemShut {NoStop}%
\bibitem [{\citenamefont {Grimme}, \citenamefont {Ehrlich},\ and\ \citenamefont
  {Goerigk}(2011)}]{grimme2011effect}%
  \BibitemOpen
  \bibfield  {author} {\bibinfo {author} {\bibfnamefont {S.}~\bibnamefont
  {Grimme}}, \bibinfo {author} {\bibfnamefont {S.}~\bibnamefont {Ehrlich}}, \
  and\ \bibinfo {author} {\bibfnamefont {L.}~\bibnamefont {Goerigk}},\
  }\bibfield  {title} {\enquote {\bibinfo {title} {{E}ffect of the {D}amping
  {F}unction in {D}ispersion {C}orrected {D}ensity {F}unctional {T}heory},}\
  }\href@noop {} {\bibfield  {journal} {\bibinfo  {journal} {J. Comput. Chem.}\
  }\textbf {\bibinfo {volume} {32}},\ \bibinfo {pages} {1456--1465} (\bibinfo
  {year} {2011})}\BibitemShut {NoStop}%
\bibitem [{\citenamefont {Kirchner}, \citenamefont {Bl{\"a}ser},\ and\
  \citenamefont {Boese}(2010)}]{kirchner2010co}%
  \BibitemOpen
  \bibfield  {author} {\bibinfo {author} {\bibfnamefont {M.~T.}\ \bibnamefont
  {Kirchner}}, \bibinfo {author} {\bibfnamefont {D.}~\bibnamefont
  {Bl{\"a}ser}}, \ and\ \bibinfo {author} {\bibfnamefont {R.}~\bibnamefont
  {Boese}},\ }\bibfield  {title} {\enquote {\bibinfo {title} {{C}o-crystals
  with {A}cetylene: {S}mall {I}s {N}ot {S}imple!}}\ }\href@noop {} {\bibfield
  {journal} {\bibinfo  {journal} {Chem.-Eur. J.}\ }\textbf {\bibinfo {volume}
  {16}},\ \bibinfo {pages} {2131--2146} (\bibinfo {year} {2010})}\BibitemShut
  {NoStop}%
\bibitem [{\citenamefont {Bussi}, \citenamefont {Donadio},\ and\ \citenamefont
  {Parrinello}(2007)}]{bussi2007canonical}%
  \BibitemOpen
  \bibfield  {author} {\bibinfo {author} {\bibfnamefont {G.}~\bibnamefont
  {Bussi}}, \bibinfo {author} {\bibfnamefont {D.}~\bibnamefont {Donadio}}, \
  and\ \bibinfo {author} {\bibfnamefont {M.}~\bibnamefont {Parrinello}},\
  }\bibfield  {title} {\enquote {\bibinfo {title} {{C}anonical {S}ampling
  through {V}elocity {R}escaling},}\ }\href@noop {} {\bibfield  {journal}
  {\bibinfo  {journal} {J. Chem. Phys.}\ }\textbf {\bibinfo {volume} {126}},\
  \bibinfo {pages} {014101} (\bibinfo {year} {2007})}\BibitemShut {NoStop}%
\bibitem [{\citenamefont {Kumar}, \citenamefont {Schmidt},\ and\ \citenamefont
  {Skinner}(2007)}]{kumar2007hydrogen}%
  \BibitemOpen
  \bibfield  {author} {\bibinfo {author} {\bibfnamefont {R.}~\bibnamefont
  {Kumar}}, \bibinfo {author} {\bibfnamefont {J.}~\bibnamefont {Schmidt}}, \
  and\ \bibinfo {author} {\bibfnamefont {J.}~\bibnamefont {Skinner}},\
  }\bibfield  {title} {\enquote {\bibinfo {title} {Hydrogen bonding definitions
  and dynamics in liquid water},}\ }\href@noop {} {\bibfield  {journal}
  {\bibinfo  {journal} {The Journal of chemical physics}\ }\textbf {\bibinfo
  {volume} {126}} (\bibinfo {year} {2007})}\BibitemShut {NoStop}%
\bibitem [{\citenamefont {Luzar}\ and\ \citenamefont
  {Chandler}(1996)}]{luzar1996hydrogen}%
  \BibitemOpen
  \bibfield  {author} {\bibinfo {author} {\bibfnamefont {A.}~\bibnamefont
  {Luzar}}\ and\ \bibinfo {author} {\bibfnamefont {D.}~\bibnamefont
  {Chandler}},\ }\bibfield  {title} {\enquote {\bibinfo {title}
  {{H}ydrogen-bond {K}inetics in {L}iquid {W}ater},}\ }\href@noop {} {\bibfield
   {journal} {\bibinfo  {journal} {Nature}\ }\textbf {\bibinfo {volume}
  {379}},\ \bibinfo {pages} {55--57} (\bibinfo {year} {1996})}\BibitemShut
  {NoStop}%
\bibitem [{\citenamefont {Markland}\ and\ \citenamefont
  {Ceriotti}(2018)}]{markland2018nuclear}%
  \BibitemOpen
  \bibfield  {author} {\bibinfo {author} {\bibfnamefont {T.~E.}\ \bibnamefont
  {Markland}}\ and\ \bibinfo {author} {\bibfnamefont {M.}~\bibnamefont
  {Ceriotti}},\ }\bibfield  {title} {\enquote {\bibinfo {title} {Nuclear
  quantum effects enter the mainstream},}\ }\href@noop {} {\bibfield  {journal}
  {\bibinfo  {journal} {Nature Reviews Chemistry}\ }\textbf {\bibinfo {volume}
  {2}},\ \bibinfo {pages} {0109} (\bibinfo {year} {2018})}\BibitemShut
  {NoStop}%
\bibitem [{\citenamefont {Gordon}(1965{\natexlab{a}})}]{gordon1965relations}%
  \BibitemOpen
  \bibfield  {author} {\bibinfo {author} {\bibfnamefont {R.}~\bibnamefont
  {Gordon}},\ }\bibfield  {title} {\enquote {\bibinfo {title} {{R}elations
  between {R}aman {S}pectroscopy and {N}uclear {S}pin {R}elaxation},}\
  }\href@noop {} {\bibfield  {journal} {\bibinfo  {journal} {J. Chem. Phys.}\
  }\textbf {\bibinfo {volume} {42}},\ \bibinfo {pages} {3658--3665} (\bibinfo
  {year} {1965}{\natexlab{a}})}\BibitemShut {NoStop}%
\bibitem [{\citenamefont {Gordon}(1965{\natexlab{b}})}]{gordon1965molecular}%
  \BibitemOpen
  \bibfield  {author} {\bibinfo {author} {\bibfnamefont {R.}~\bibnamefont
  {Gordon}},\ }\bibfield  {title} {\enquote {\bibinfo {title} {{M}olecular
  {M}otion in {I}nfrared and {R}aman {S}pectra},}\ }\href@noop {} {\bibfield
  {journal} {\bibinfo  {journal} {J. Chem. Phys.}\ }\textbf {\bibinfo {volume}
  {43}},\ \bibinfo {pages} {1307--1312} (\bibinfo {year}
  {1965}{\natexlab{b}})}\BibitemShut {NoStop}%
\bibitem [{\citenamefont {Thakur}\ and\ \citenamefont
  {Remsing}(2021)}]{thakur2021distributed}%
  \BibitemOpen
  \bibfield  {author} {\bibinfo {author} {\bibfnamefont {A.~C.}\ \bibnamefont
  {Thakur}}\ and\ \bibinfo {author} {\bibfnamefont {R.~C.}\ \bibnamefont
  {Remsing}},\ }\bibfield  {title} {\enquote {\bibinfo {title} {Distributed
  charge models of liquid methane and ethane for dielectric effects and
  solvation},}\ }\href@noop {} {\bibfield  {journal} {\bibinfo  {journal}
  {Molecular Physics}\ }\textbf {\bibinfo {volume} {119}},\ \bibinfo {pages}
  {e1933228} (\bibinfo {year} {2021})}\BibitemShut {NoStop}%
\bibitem [{\citenamefont {Lipari}\ and\ \citenamefont
  {Szabo}(1981)}]{lipari1981pade}%
  \BibitemOpen
  \bibfield  {author} {\bibinfo {author} {\bibfnamefont {G.}~\bibnamefont
  {Lipari}}\ and\ \bibinfo {author} {\bibfnamefont {A.}~\bibnamefont {Szabo}},\
  }\bibfield  {title} {\enquote {\bibinfo {title} {Pade approximants to
  correlation functions for restricted rotational diffusion},}\ }\href@noop {}
  {\bibfield  {journal} {\bibinfo  {journal} {The Journal of Chemical Physics}\
  }\textbf {\bibinfo {volume} {75}},\ \bibinfo {pages} {2971--2976} (\bibinfo
  {year} {1981})}\BibitemShut {NoStop}%
\bibitem [{\citenamefont {Lipari}\ and\ \citenamefont
  {Szabo}(1980)}]{lipari1980effect}%
  \BibitemOpen
  \bibfield  {author} {\bibinfo {author} {\bibfnamefont {G.}~\bibnamefont
  {Lipari}}\ and\ \bibinfo {author} {\bibfnamefont {A.}~\bibnamefont {Szabo}},\
  }\bibfield  {title} {\enquote {\bibinfo {title} {Effect of librational motion
  on fluorescence depolarization and nuclear magnetic resonance relaxation in
  macromolecules and membranes},}\ }\href@noop {} {\bibfield  {journal}
  {\bibinfo  {journal} {Biophysical journal}\ }\textbf {\bibinfo {volume}
  {30}},\ \bibinfo {pages} {489--506} (\bibinfo {year} {1980})}\BibitemShut
  {NoStop}%
\bibitem [{\citenamefont {Neish}\ \emph {et~al.}(2015)\citenamefont {Neish},
  \citenamefont {Barnes}, \citenamefont {Sotin}, \citenamefont {MacKenzie},
  \citenamefont {Soderblom}, \citenamefont {Le~Mou{\'e}lic}, \citenamefont
  {Kirk}, \citenamefont {Stiles}, \citenamefont {Malaska}, \citenamefont
  {Le~Gall}, \citenamefont {Brown}, \citenamefont {Baines}, \citenamefont
  {Buratti}, \citenamefont {Clark},\ and\ \citenamefont
  {Nicholson}}]{NeishGRL2015}%
  \BibitemOpen
  \bibfield  {author} {\bibinfo {author} {\bibfnamefont {C.~D.}\ \bibnamefont
  {Neish}}, \bibinfo {author} {\bibfnamefont {J.~W.}\ \bibnamefont {Barnes}},
  \bibinfo {author} {\bibfnamefont {C.}~\bibnamefont {Sotin}}, \bibinfo
  {author} {\bibfnamefont {S.}~\bibnamefont {MacKenzie}}, \bibinfo {author}
  {\bibfnamefont {J.~M.}\ \bibnamefont {Soderblom}}, \bibinfo {author}
  {\bibfnamefont {S.}~\bibnamefont {Le~Mou{\'e}lic}}, \bibinfo {author}
  {\bibfnamefont {R.~L.}\ \bibnamefont {Kirk}}, \bibinfo {author}
  {\bibfnamefont {B.~W.}\ \bibnamefont {Stiles}}, \bibinfo {author}
  {\bibfnamefont {M.~J.}\ \bibnamefont {Malaska}}, \bibinfo {author}
  {\bibfnamefont {A.}~\bibnamefont {Le~Gall}}, \bibinfo {author} {\bibfnamefont
  {R.~H.}\ \bibnamefont {Brown}}, \bibinfo {author} {\bibfnamefont {K.~H.}\
  \bibnamefont {Baines}}, \bibinfo {author} {\bibfnamefont {B.}~\bibnamefont
  {Buratti}}, \bibinfo {author} {\bibfnamefont {R.~N.}\ \bibnamefont {Clark}},
  \ and\ \bibinfo {author} {\bibfnamefont {P.~D.}\ \bibnamefont {Nicholson}},\
  }\bibfield  {title} {\enquote {\bibinfo {title} {Spectral properties of
  titan's impact craters imply chemical weathering of its surface},}\
  }\href@noop {} {\bibfield  {journal} {\bibinfo  {journal} {Geophysical
  Research Letters}\ }\textbf {\bibinfo {volume} {42}},\ \bibinfo {pages}
  {3746--3754} (\bibinfo {year} {2015})}\BibitemShut {NoStop}%
\bibitem [{\citenamefont {Yu}\ \emph {et~al.}(2018)\citenamefont {Yu},
  \citenamefont {H{\"o}rst}, \citenamefont {He}, \citenamefont {McGuiggan},\
  and\ \citenamefont {Crawford}}]{yu2018does}%
  \BibitemOpen
  \bibfield  {author} {\bibinfo {author} {\bibfnamefont {X.}~\bibnamefont
  {Yu}}, \bibinfo {author} {\bibfnamefont {S.~M.}\ \bibnamefont {H{\"o}rst}},
  \bibinfo {author} {\bibfnamefont {C.}~\bibnamefont {He}}, \bibinfo {author}
  {\bibfnamefont {P.}~\bibnamefont {McGuiggan}}, \ and\ \bibinfo {author}
  {\bibfnamefont {B.}~\bibnamefont {Crawford}},\ }\bibfield  {title} {\enquote
  {\bibinfo {title} {Where does titan sand come from: insight from mechanical
  properties of titan sand candidates},}\ }\href@noop {} {\bibfield  {journal}
  {\bibinfo  {journal} {Journal of Geophysical Research: Planets}\ }\textbf
  {\bibinfo {volume} {123}},\ \bibinfo {pages} {2310--2321} (\bibinfo {year}
  {2018})}\BibitemShut {NoStop}%
\bibitem [{\citenamefont {Solomonidou}\ \emph {et~al.}(2020)\citenamefont
  {Solomonidou}, \citenamefont {Neish}, \citenamefont {Coustenis},
  \citenamefont {Malaska}, \citenamefont {Le~Gall}, \citenamefont {Lopes},
  \citenamefont {Werynski}, \citenamefont {Markonis}, \citenamefont {Lawrence},
  \citenamefont {Altobelli} \emph {et~al.}}]{solomonidou2020chemical}%
  \BibitemOpen
  \bibfield  {author} {\bibinfo {author} {\bibfnamefont {A.}~\bibnamefont
  {Solomonidou}}, \bibinfo {author} {\bibfnamefont {C.}~\bibnamefont {Neish}},
  \bibinfo {author} {\bibfnamefont {A.}~\bibnamefont {Coustenis}}, \bibinfo
  {author} {\bibfnamefont {M.}~\bibnamefont {Malaska}}, \bibinfo {author}
  {\bibfnamefont {A.}~\bibnamefont {Le~Gall}}, \bibinfo {author} {\bibfnamefont
  {R.~M.}\ \bibnamefont {Lopes}}, \bibinfo {author} {\bibfnamefont
  {A.}~\bibnamefont {Werynski}}, \bibinfo {author} {\bibfnamefont
  {Y.}~\bibnamefont {Markonis}}, \bibinfo {author} {\bibfnamefont
  {K.}~\bibnamefont {Lawrence}}, \bibinfo {author} {\bibfnamefont
  {N.}~\bibnamefont {Altobelli}},  \emph {et~al.},\ }\bibfield  {title}
  {\enquote {\bibinfo {title} {The chemical composition of impact craters on
  titan-i. implications for exogenic processing},}\ }\href@noop {} {\bibfield
  {journal} {\bibinfo  {journal} {Astronomy \& Astrophysics}\ }\textbf
  {\bibinfo {volume} {641}},\ \bibinfo {pages} {A16} (\bibinfo {year}
  {2020})}\BibitemShut {NoStop}%
\bibitem [{\citenamefont {Yu}\ \emph {et~al.}(2023)\citenamefont {Yu},
  \citenamefont {Yu}, \citenamefont {Garver}, \citenamefont {Li}, \citenamefont
  {Hawthorn}, \citenamefont {Sciamma-O'Brien}, \citenamefont {Zhang},\ and\
  \citenamefont {Barth}}]{yu2023material}%
  \BibitemOpen
  \bibfield  {author} {\bibinfo {author} {\bibfnamefont {X.}~\bibnamefont
  {Yu}}, \bibinfo {author} {\bibfnamefont {Y.}~\bibnamefont {Yu}}, \bibinfo
  {author} {\bibfnamefont {J.}~\bibnamefont {Garver}}, \bibinfo {author}
  {\bibfnamefont {J.}~\bibnamefont {Li}}, \bibinfo {author} {\bibfnamefont
  {A.}~\bibnamefont {Hawthorn}}, \bibinfo {author} {\bibfnamefont
  {E.}~\bibnamefont {Sciamma-O'Brien}}, \bibinfo {author} {\bibfnamefont
  {X.}~\bibnamefont {Zhang}}, \ and\ \bibinfo {author} {\bibfnamefont
  {E.}~\bibnamefont {Barth}},\ }\bibfield  {title} {\enquote {\bibinfo {title}
  {Material properties of organic liquids, ices, and hazes on titan},}\
  }\href@noop {} {\bibfield  {journal} {\bibinfo  {journal} {The Astrophysical
  Journal Supplement Series}\ }\textbf {\bibinfo {volume} {266}},\ \bibinfo
  {pages} {30} (\bibinfo {year} {2023})}\BibitemShut {NoStop}%
\bibitem [{\citenamefont {Towns}\ \emph {et~al.}(2014)\citenamefont {Towns},
  \citenamefont {Cockerill}, \citenamefont {Dahan}, \citenamefont {Foster},
  \citenamefont {Gaither}, \citenamefont {Grimshaw}, \citenamefont {Hazlewood},
  \citenamefont {Lathrop}, \citenamefont {Lifka}, \citenamefont {Peterson},
  \citenamefont {Roskies}, \citenamefont {Scott},\ and\ \citenamefont
  {Wilkins-Diehr}}]{towns2014xsede}%
  \BibitemOpen
  \bibfield  {author} {\bibinfo {author} {\bibfnamefont {J.}~\bibnamefont
  {Towns}}, \bibinfo {author} {\bibfnamefont {T.}~\bibnamefont {Cockerill}},
  \bibinfo {author} {\bibfnamefont {M.}~\bibnamefont {Dahan}}, \bibinfo
  {author} {\bibfnamefont {I.}~\bibnamefont {Foster}}, \bibinfo {author}
  {\bibfnamefont {K.}~\bibnamefont {Gaither}}, \bibinfo {author} {\bibfnamefont
  {A.}~\bibnamefont {Grimshaw}}, \bibinfo {author} {\bibfnamefont
  {V.}~\bibnamefont {Hazlewood}}, \bibinfo {author} {\bibfnamefont
  {S.}~\bibnamefont {Lathrop}}, \bibinfo {author} {\bibfnamefont
  {D.}~\bibnamefont {Lifka}}, \bibinfo {author} {\bibfnamefont {G.~D.}\
  \bibnamefont {Peterson}}, \bibinfo {author} {\bibfnamefont {R.}~\bibnamefont
  {Roskies}}, \bibinfo {author} {\bibfnamefont {J.~R.}\ \bibnamefont {Scott}},
  \ and\ \bibinfo {author} {\bibfnamefont {N.}~\bibnamefont {Wilkins-Diehr}},\
  }\bibfield  {title} {\enquote {\bibinfo {title} {{XSEDE}: {A}ccelerating
  {S}cientific {D}iscovery},}\ }\href@noop {} {\bibfield  {journal} {\bibinfo
  {journal} {Comput. Sci. Eng.}\ }\textbf {\bibinfo {volume} {16}},\ \bibinfo
  {pages} {62--74} (\bibinfo {year} {2014})}\BibitemShut {NoStop}%
\end{thebibliography}%

\end{document}